\documentstyle[preprint,aps]{revtex}
\tightenlines
\input psfig.sty
\begin{document}
\title{Melosh Construction Of Relativistic Three-Quark Baryon Wave Functions}
\author{Xuepeng Sun and H. J. Weber}
\address{Institute of Nuclear and Particle Physics and Department of Physics,\\
University of Virginia,\\Charlottesville VA 22904-4714, USA}
\maketitle
\begin{abstract}
A method for constructing a {\bf complete set} of relativistic three-quark 
states in light front dynamics is implemented for the 
nucleon, N$^*$(1520) and N$^*$(1535). This approach facilitates constructing 
states containing virtual antiquarks and a physical interpretation is 
provided in terms of transition amplitudes from quark to quark-gluon or 
quark-Goldstone boson Fock states of chiral dynamics generated by flux tube 
breaking expected in QCD at intermediate distances.   

\end{abstract}
\indent Bargmann-Wigner three-quark bases.  
\newpage   

\section{Introduction}
In light front dynamics~\cite{lf} (LF) and in Dirac's point form of 
relativistic few-body physics~\cite{pamd},\cite{graz}, hadron wave functions 
may be boosted kinematically, that is, independent of interactions. Because 
form factors depend on boosted wave functions this attractive feature has 
motivated many recent electroweak form factor calculations~\cite{lcqm}.  

Because there is no spontaneous creation of massive fermions in LF 
dynamics that is quantized on the null plane, the quantized vacuum is fairly 
simple, so that a constituent Fock space expansion is practical where partons 
are directly related to the hadron.  

Here we construct relativistic baryon basis states in light front 
dynamics, although all important results are also valid in the point form, 
where four-velocities are replacing the light cone momentum variables. The 
null plane is invariant under seven of the ten Poincar\'e generators. A 
subgroup of six of these generators form the stability group that acts 
transitively on states of a given mass-shell hyperboloid. As a result, the 
total momentum separates from the internal momentum variables~\cite{lst}. 
This is one of the reasons why LF dynamics is usually formulated in momentum 
space. Thus, wave functions depend only on the relative momentum variables 
and, being invariant under kinematic Poincar\'e transformations, are completely 
determined once they are known at rest.      

Except for Sect. III we deal with nucleon basis states. Except for Sect. V, we 
have applications in mind to spacelike electroweak form factors in the Breit 
frame (with momentum transfer $q^+=q^0+q^3=0$) of LF dynamics excluding the 
complications of timelike exclusive processes (which can be 
treated~\cite{dgns}). As a rule, applications involve truncating the 
Fock space expansion by particle number, which does not violate Lorentz 
invariance because in LF dynamics the boost operators are kinematic, that 
is, do not contain interactions.      

Three-quark wave functions for the nucleon have been constructed in the 
constituent quark model as products of a totally symmetric momentum and a 
nonstatic spin-flavor wave function which is an eigenstate of the spin and 
total angular momentum (squared) and its projection on the light cone axis.
When nonstatic spin-flavor wave functions are built from the nonrelativistic
quark model (NQM) via Melosh rotations~\cite{m}, (see ref.~\cite{bkw} for a 
review and refs.) they form a Hilbert space of relativistic three-quark states 
that we shall call the Pauli-Melosh basis. It is in one-to-one correspondence 
with the NQM and manifestly orthogonal.  

An alternative effective field theory approach starts from three-quark nucleon 
interaction Lagrangians~\cite{fbw} in light front dynamics which provide the 
nonstatic spin-flavor wave functions in a Dirac matrix representation. When 
the triangle Feynman diagram for a form factor is projected to the null-plane 
the radial momentum wave functions are defined in terms of three-quark-nucleon 
vertex functions and a totally symmetric energy denominator (three-quark 
propagator). Based on Lorentz covariance, parity and SU(2) isospin invariance 
there are eight independent couplings and three totally symmetric couplings.   
This Dirac basis is manifestly Lorentz covariant under kinematic 
transformations.    

Here we wish to address the problem in that there are significantly more 
relativistic hadron wave functions in the Dirac basis than compared to the 
Pauli-Melosh basis. For example, there are three nucleon states compared  
to the single S-state of the NQM in the static limit and five N$^*$(1535) 
states. 

The alternative Bargmann-Wigner basis~\cite{bw} (BW), which is in one-to-one 
correspondence with the Dirac basis~\cite{bkw}, sheds light on this 
problem from another point of view. It is based on the equivalence of the 
infinite momentum frame (IMF) and light front dynamics which implies that 
quarks bound in a hadron in the IMF are all collinear, that 
is, have equal velocities $p_i/m=P/M,$ where $M$ is the baryon mass and 
$m$ a constituent quark mass. The BW basis constructs three-quark baryon 
Fock states in terms of baryon Dirac spinors ($U, V$). There is only one 
totally symmetric nucleon basis state ($UUU$). It corresponds to the nucleon 
ground state of the Pauli-Melosh basis. The other two totally symmetric 
three-quark states of the BW and Dirac bases contain two Dirac $V$-spinors and 
one $U$-spinor. 

In this paper it is shown that a set of spin-rotated spinors are required to 
account for all three-quark-nucleon couplings, that is, all three non-static 
spin-flavor wave functions of the nucleon. Corresponding results are valid for 
N$^*$s, the baryon octet and other baryons. 

The paper is organized as follows. In Sect. II we introduce spin-rotated 
spinors based on the complete $U,~V$ basis, which are used in Sect. IV to 
construct the complete basis of three-quark spin-flavor wave functions in LF 
dynamics. In Sect. III we provide the transformation between the BW and Dirac 
representations of the three-quark couplings for the nucleon, N$^*$(1520) and 
N$^*$(1535) in a transparent approach that avoids overlap matrix elements of 
both bases used in ref~\cite{bkw}. After symmetrizing we find three and five 
spin-flavor components for nucleon and N$^*$(1520),~N$^*$(1535), respectively. 
In Sect. IV we start from the instant form to construct the complete set of 
relativistic spin-flavor wave functions for positive-energy quarks. In Sect. V 
we consider the possibility of $v$-spinors in the wave functions and discuss 
their possible origin in light front dynamics. We discuss a physical 
interpretation of baryon states with $v$ spinors as part of transition 
amplitudes from three-quark states to quark-gluon or quark-Goldstone boson 
Fock states via flux-tube breaking at intermediate distances in QCD.

\section{Spin-Rotated Light-Cone Spinors}

When the bound quarks of the nucleon are in the infinite momentum frame (IMF), 
or equivalently in the null plane of light front dynamics (LF), they are all 
collinear~\cite{hk} 
\begin{eqnarray}
\frac{p_i}{m_i}=\frac{P}{M}, \rm{~as~} P\to \infty.
\end{eqnarray}   
The transformation to the IMF amounts to a change of the 
usual momentum variables~\cite{ls}  
\begin{eqnarray}
p^{\mu}=(p^0,{\bf p}) \rm{~to~} (p^+=p^0+p^3,~ {\bf p}_{\perp}=(p^1, p^2),~ 
p^-=p^0-p^3) 
\end{eqnarray}
to light cone momentum components. As in any Hamiltonian version of field 
theory, partons are on their mass shell so that the light cone energy 
$p^-=(m^2+{\bf p}^2_{\perp})/p^+>0,$ in contrast to the square root ambiguity 
for $p^0$ in the instant form. In transitions the $p^-$ variable is not 
conserved. 

The usual instant-form quark spinors~\cite{bd} $u_{\lambda}^{inst}(k),~
v_{\lambda}^{inst}(k)$ under this transformation to the IMF will change their 
form and become light-cone spinors $u_{\lambda}^{LC}(k),~v_{\lambda}^{LC}(k)$ 
denoted by a superscript $LC$ (see ref.~\cite{bkw} for a review) and 
$\lambda$ is the helicity.  Note that 
our $v$-spinors obey $v_{\lambda}(k)=\gamma_5 u_{\lambda}(k)$ in contrast to 
$v_{\lambda}=C\bar u_{\lambda}^T$ of ref.~\cite{bd}.   

The total momentum spinors $U_{\uparrow}$,~$U_{\downarrow}$ of the nucleon 
satisfy the free Dirac equation
\begin{eqnarray}
(\gamma \cdot P - M) U_{\lambda}=0,\quad 
\label{dir}
\end{eqnarray}  
where $M$ is the nucleon mass, $P$ its total momentum. In the rest frame of 
the nucleon the total momentum spinors have the form 
$U^T_{\uparrow}=(1,0,0,0),~U^T_{\downarrow}=(0,1,0,0).$ The corresponding 
$V_{\uparrow}$,~$V_{\downarrow}$ spinors satisfy 
\begin{eqnarray}
(\gamma \cdot P+M)V_{\lambda}=0, 
\end{eqnarray}
and have the rest-frame forms $V^T_{\uparrow}=(0,0,1,0),~V_{\downarrow}
=(0,0,0,1).$ The $U,~V$ spinors are the building blocks of the Bargmann-Wigner 
(BW) basis. 

The unitary transformation between instant-form and light front spinors is 
given by the Melosh rotation matrices~\cite{as}  
\begin{equation}
u_{\lambda}^{inst}=\sum_{\xi}{\cal R}_{\lambda\xi}^{(1)}u_{\xi}^{LC},\quad
v_{\lambda}^{inst}=\sum_{\xi}{\cal R}_{\lambda\xi}^{(2)}v_{\xi}^{LC},
\label{trf}
\end{equation}
where $\lambda$ and $\xi$ represent the helicities and 
\begin{equation}
{\cal R}_{\lambda\xi}^{(1)}={\cal R}_{\lambda\xi}^{(2)}={1\over {\sqrt{2k^+ 
(m+k^0)}}}{\left(\begin{array}{cc}k^++m & -k^R\\
k^L & k^++m \end{array}\right)}_{\lambda\xi}.
\label{melosh}
\end{equation}
Here $k^{R,L}=k^1\pm ik^2$ are conventional abbreviations for the quark 
momentum in the nucleon rest frame and $m$ is the quark mass.
The Melosh matrix for a single quark can also be written as an overlap of a  
quark spinor and the nucleon spinors $U_{\lambda},~V_{\lambda},$ 
\begin{eqnarray}
\bar u_{\xi}^{LC}U_{\lambda}={1\over 2{\sqrt{m k^+}}}
{\left(\begin{array}{cc}k^++m & -k^R\\k^L & k^++m \end{array}\right)}_{\lambda
\xi}\\
\bar v_{\xi}^{LC}V_{\lambda}=-{1\over 2{\sqrt{m k^+}}}{\left(\begin{array}
{cc}k^++m & -k^R\\k^L & k^++m \end{array}\right)}_{\lambda\xi}
\end{eqnarray}
in the nucleon rest frame, so that the Melosh rotations may be written in the 
shorter form 
\begin{eqnarray}
{\cal R}_{\lambda\xi}^{(1)}=N\bar u_{\xi}^{LC}U_{\lambda}={\cal R}_{\lambda\xi}
^{(2)}=-N\bar v_{\xi}^{LC}V_{\lambda}, 
\label{rtrf}
\end{eqnarray}
with normalization $N=\sqrt{2m\over{m+k^0}}$. In any frame, 
therefore,  
\begin{eqnarray}
\left(\begin{array}{cc}\bar u_{\uparrow}^{LC}U_{\uparrow} & 
\bar u_{\downarrow}^{LC}U_{\uparrow}\\ 
\bar u_{\uparrow}^{LC}U_{\downarrow} & \bar u_{\downarrow}^{LC}U_{\downarrow}\\
\end{array}\right)=
\frac{1}{2\sqrt{mMp^+P^+}}\left(\begin{array}{cc} Mp^++mP^+ & P^Rp^+-P^+p^R\\ 
P^+p^L-P^Lp^+ & Mp^++mP^+\\ 
\end{array}\right). 
\end{eqnarray}
In a different notation these Melosh rotations are considered in~\cite{as}. 

The block diagonal form of the transformations, Eq.~\ref{rtrf}, motivates us 
to define off-diagonal transformations 
\begin{eqnarray}
w_{\lambda}^{inst}=\sum_l \left(\bar v_l^{LC}U_{\lambda}\right)u_l^{LC},\quad
z_{\lambda}^{inst}=\sum_l \left(\bar u_l^{LC}V_{\lambda}\right)v_l^{LC},  
\label{4matrix2}
\end{eqnarray}
so that   
\begin{eqnarray}
\bar v_{\xi}^{LC}U_{\lambda}&=&{1\over 2{\sqrt{m p^+}}}{\left(\begin{array}{cc}
k^+-m & k^R\\k^L & -k^++m \end{array}\right)}_{\lambda\xi}\equiv 
{1\over N'}{\cal R}_{\lambda\xi}^{(3)}\\
\bar u_{\xi}^{LC}V_{\lambda}&=&-{1\over 2{\sqrt{m p^+}}}{\left(\begin{array}{cc}
k^+-m & k^R\\k^L & -k^++m \end{array}\right)}_{\lambda\xi}\equiv 
-{1\over N'}{\cal R}_{\lambda\xi}^{(4)},
\end{eqnarray}
with normalization $N'\equiv\sqrt {2m\over k^0-m}$. These transformations are 
manifestly unitary and relate  
\begin{eqnarray}
w_{\lambda}^{inst}=\sum_{\xi}{\cal R}_{\lambda\xi}^{(3)}v_{\xi}^{LC},\quad 
z_{\lambda}^{inst}=\sum_{\xi}{\cal R}_{\lambda\xi}^{(4)}u_{\xi}^{LC}.
\label{wztrf}
\end{eqnarray}

The $z^{inst}_{\lambda},~w^{inst}_{\lambda}$ spinors obey the Dirac equations 
\begin{eqnarray}
(\gamma \cdot k-m)z^{inst}_{\lambda}(k)=0,\quad (\gamma \cdot k+m)
w^{inst}_{\lambda}(k)=0.   
\end{eqnarray}
We can see that the $w^{inst}$ and $z^{inst}$ spinors differ from $u^{inst}$ 
and $v^{inst}$ spinors only in replacing $k^0+m$ by $k^0-m$. Thus in highly 
relativistic cases, $k^0+m\sim k^0-m$ and $w_{\lambda}^{inst}\rightarrow 
u_{\lambda}^{inst}$, $z_{\lambda}^{inst} \rightarrow v_{\lambda}^{inst}.$ 
In the static limit, however, the $z^{inst},~w^{inst}$ spinors appear to 
diverge, but on closer inspection the static limit is finite yet depends on 
the direction in which it is taken. Let us approach the static limit so that
\begin{eqnarray} 
k^x=k \alpha,~k^y=k \beta,~k^z=k \gamma,\quad \alpha^2+\beta^2+\gamma^2=1,~
\rm{~as~} k\to 0.  
\end{eqnarray} 
Then 
\begin{eqnarray}
\sqrt{k^0-m}\sim \frac{k}{\sqrt{2m}},~ \frac{k^z}{\sqrt{k^0-m}}\sim \gamma 
\sqrt{2m},~ \frac{k^R}{\sqrt{k^0-m}}\sim (\alpha+i\beta)\sqrt{2m} 
\rm{~as~} k\to 0, 
\end{eqnarray}
and $$(z_{\uparrow}^{inst})^T= (\gamma,~\alpha+i\beta,~0,~0)$$ remains finite 
and normalized to unity, but clearly depends on how one approaches the 
static limit. Similar results hold for the other three spinors. 

In fact, the $w^{inst},~z^{inst}$ spinors are directly related to the 
$u^{inst},~v^{inst}$ spinors by the spin rotation 
\begin{eqnarray}
w_{\lambda}^{inst}(k)=\vec{\sigma}\cdot \hat{k} v_{\lambda}^{inst}(k),\quad 
z_{\lambda}^{inst}(k)=\vec{\sigma}\cdot \hat{k} u_{\lambda}^{inst}(k),    
\end{eqnarray}
where the unit vector $\hat{k}=\vec{k}/k,~k=\sqrt{\vec{k}^2}$. This spin 
rotation is clearly not well defined for $k\to 0$.
To prove this relation for $z^{inst}$, for example, we repeatedly substitute  
\begin{equation}
\vec{k}^2=(k^0)^2-m^2=(k^0-m)(k^0+m)
\end{equation}
in
\begin{eqnarray}\nonumber
\vec{\sigma}\cdot \hat{k} u^{inst}(k)&=&\sqrt {{k^0+m} \over 2m} 
\pmatrix{\vec{\sigma}\cdot \hat{k}\cr \frac{k}{k^0+m}}\chi
=\frac{k}{\sqrt{2m(k^0+m)}}\pmatrix{\frac{k^0+m}{k^2}\vec{\sigma}\cdot \vec{k}
\cr 1}\chi\\&=&\sqrt{{k^0-m}\over 2m}\pmatrix{\frac{\vec{\sigma}\cdot \vec{k}}
{k^0-m}\cr 1}\chi =z^{inst}.  
\end{eqnarray}         
The consistency of this ambiguity of the static limit with the nature of the 
$UVV$ invariants is discussed in Sect. IV.   
\par
The reason for introducing these spin rotated $w$ and $z$ spinors and their 
usefulness become transparent for multi-quark states only. But let us 
emphasize here already that we may view the matrix elements involved in 
Eq.~\ref{4matrix2} as unitary transformations from light cone to instant 
spinors via the BW (or IMF) baryon basis states $U,~V$. We shall show that, 
for the resulting {\bf three-quark} basis to be complete, one needs to go via 
$V$ states as well as $U$ states. In this sense Eq.~\ref{4matrix2} provides 
the missing off-diagonal elements from $U_{\lambda}$ to $v_{\lambda'}$ and 
$V_{\lambda}$ to $u_{\lambda'}$.    

In Section IV we shall use these Melosh rotations for $w$ and $z$ states to 
construct all three-quark states of the Dirac and BW 
representations, which will make their equivalence manifest. 

\section{Conversion of Dirac to Bargmann-Wigner Basis}
In this section we review the connection between two forms of nucleon wave 
functions, one in the Dirac basis and the other in the Bargmann-Wigner basis. 
We develop a simple and transparent method to establish the one-to-one 
correspondence between the Dirac and Bargmann-Wigner bases that avoids the 
evaluation of overlap matrix elements used in ref.~\cite{bkw}. 
  
When the nucleon is treated as a spin-$\frac{1}{2}$ field, it is a third-rank 
spinor $\Psi_{[\alpha \beta]\gamma}$ which satisfies 
the free Dirac equation for each spinor index
\begin{eqnarray}
(\gamma \cdot P - M)^{\alpha'}_{\alpha}\Psi_{[\alpha' \beta]\gamma}= 
(\gamma \cdot P - M)^{\beta'}_{\beta}\Psi_{[\alpha \beta']\gamma}=
(\gamma \cdot P - M)^{\gamma'}_{\gamma}\Psi_{[\alpha \beta]\gamma'} =0.  
\label{bdir}
\end{eqnarray}
These constraints and the total symmetry under permutation of the three spinor 
indices lead to the spinor form $[(\gamma \cdot P+M)\gamma_5 C]_{\alpha 
\beta}U_{\lambda}(P)$ of $\Psi_{[\alpha \beta]\gamma}$~\cite{bw}, which is 
antisymmetric under the exchange of the $\alpha, \beta$ indices. Here 
$C=i\gamma^2\gamma^0$ is the charge conjugation matrix. 

However, in QCD and in quark models in particular the nucleon is no longer a 
local field. Various bound state nucleon wave function components may be 
related to different three-quark-nucleon vertices defined by corresponding 
interaction Lagrangians. Such an approach has recently~\cite{fbw} been 
adopted for a null-plane projection of the Feynman triangle diagram for 
spacelike electroweak form factors of the nucleon. From Lorentz invariance and 
symmetries under permutations of three quarks, a basis of Dirac $\gamma$-matrix 
representations has been constructed for the spin-flavor components of these 
couplings (see ref.~~\cite{bkw} for a review) which is equivalent to the 
BW-basis. The specific spin coupling listed above reduces to the 
nonrelativistic quark model $S$-state in the static limit. This spin wave 
function dominates the applications of relativistic quark models.      

The wave function of a proton can be written in the uds basis as
\begin{equation}
\begin{array}{lllll}
|P_{\uparrow}\rangle&=&|\rm{mom.}\rangle\otimes[{1\over \sqrt 2}
(\uparrow\downarrow-\downarrow\uparrow)\uparrow \otimes {1\over\sqrt 2}
(ud-du)u\\
&+&(\rm{MS})_{spin}\otimes(\rm{MS})_{isospin}]\otimes|SU(3)_{color}\rangle,
\end{array}
\end{equation}
where MS stands for the mixed symmetric combination. 

In order to convert the spin structure of this wave function to Lorentz 
covariant form, we start rewriting two-particle spin wave functions in terms of 
Pauli matrices. We know that $\mid \uparrow \rangle \mid \downarrow \rangle 
-\mid \downarrow \rangle \mid \uparrow \rangle $ is the 
antisymmetric combination of two spin-${1\over 2}$ particles. If we 
recognize that $\mid \uparrow \rangle \rightarrow {1\choose 0}$,~ 
$\mid \downarrow \rangle \rightarrow {0 \choose 1} $ and carry out the direct 
product, then we get
\begin{eqnarray}
\label{pl1}
\bigg ({1\choose 0}{\left (0\quad 1 \right)}-{0\choose 1}{\left (1 \quad 
0\right)}\bigg)=\left( \begin{array}{cc}0 & 1\\-1 & 0 \end{array} \right) 
= -i\sigma_2,   
\end{eqnarray}
that is, the off-diagonal matrix $\sigma_2$ represents antisymmetric coupling. 
This identity has a similar form 
when we go to the four-dimensional Dirac space, where the basis is comprised of 
four-components Dirac spinors 
$U_{\uparrow},U_{\downarrow},V_{\uparrow},V_{\downarrow}$ of the nucleon from 
the previous Sect.II. For this case the decomposition takes the form 
\begin{equation}
\Gamma^{\alpha\beta}=(U^{\alpha}_{\uparrow}U^{\alpha}_{\downarrow}
V^{\alpha}_{\uparrow}V^{\alpha}_{\downarrow})_{\mu}
\Gamma^{\mu\nu}\pmatrix{U^{\beta}_{\uparrow}\cr 
U^{\beta}_{\downarrow} \cr V^{\beta}_{\uparrow} \cr 
V^{\beta}_{\downarrow}\cr}_{\nu}, 
\label{gammadef}
\end{equation}
where $\Gamma$ is one of the sixteen Dirac matrices 
\begin{eqnarray}
\gamma^{\mu} C,\sigma^{\mu\nu} C~(1\leftrightarrow 2~\rm{symmetric}),\nonumber\\ 
\gamma^{\mu}\gamma_5 C, \gamma_5 C, C~(1\leftrightarrow 2~\rm{antisymmetric}).
\label{gamex}
\end{eqnarray} 
The entries for the $\Gamma$ matrices have a twofold meaning: one as the 
coupling of the U,V spinors with different helicities and the other as the 
direct product of the spinor elements.  

The above identity~(\ref{gammadef}) is most easily checked in the nucleon rest 
frame. To see how to convert the $G_i$ of the Dirac basis in 
Table~\ref{123ndm} into $U,V$ spinors we proceed in two 
steps. First we decompose the $UV$ coupling and then the spin structure. For 
example, for $G_2 \sim \gamma_5 C \otimes U^\uparrow$ because
\begin{equation}
\begin{array}{llll}
& \gamma_5 C & = & \gamma_5 i \gamma^2 \gamma^0={\left(\begin{array}{cc}1 & 0\\
0 & 1 \end{array}\right)}^{UV} \otimes {\left(\begin{array}{cc}0 & -1\\1 & 0 
\end{array} 
\right)}^{spin}\\
& & = & {\left (U \quad V \right)}{\left(\begin{array}{cc}1 & 0\\0 & 1 
\end{array} \right)}{U \choose V} \otimes 
{\left (\uparrow \quad \downarrow \right)}{\left(\begin{array}{cc}0 & -1\\
1 & 0 \end{array} \right)}{\uparrow \choose \downarrow}\\
& & = & (UU + VV)\otimes(-\uparrow\downarrow+\downarrow\uparrow).
\end{array}
\label{g1uv1}
\end{equation}
Including the third quark, we obtain  
\begin{eqnarray} 
G_2=\gamma_5 C U_{\uparrow} &=& (UU+VV)\otimes(-\uparrow\downarrow+\downarrow
\uparrow)\otimes U^\uparrow\nonumber\\
&=& -U^\uparrow U^\downarrow U^\uparrow 
    +U^\downarrow U^\uparrow U^\uparrow 
    -V^\uparrow V^\downarrow U^\uparrow 
    +V^\downarrow V^\uparrow U^\uparrow.
\label{g1}
\end{eqnarray}
The conversion to the BW basis for all the eight invariants in 
Table~\ref{123ndm} proceeds in this way, and they are displayed in 
Table~\ref{dmbw}.

The spin part of the wave function is constructed by choosing two of the 
quarks coupling via $(\Gamma^1)_{\alpha\beta}$, abbreviated as  
$(\alpha,\beta)$, with $\Gamma^1$ from Eqs.~\ref{gammadef}, \ref{gamex} having 
definite permutation symmetry with respect to the exchange of the two spinor 
indices. The third quark index is added by combining a  
Dirac matrix $\Gamma^2$, which is one of the basic sixteen 1, $\gamma_5$, 
$\gamma^{\mu}$,$\gamma_5\gamma^{\mu}$, $\sigma_{\mu\nu}$ with the nucleon 
spinor $u_N(P)$. Hence the total spinor components have the third-rank tensor 
structure $$\Psi_{[\alpha\beta]\gamma}=(\Gamma^1)_{\alpha\beta}(\Gamma^2 
u_N(P))_{\gamma},$$ which may be evaluated for $u$- and $v$-spinors. In light 
front quark model applications, such as form factors defined by a triangle 
Feynman diagram, such spin couplings are sandwiched between $u$-spinors 
only and called spin wave functions. Next the isospin coupling is 
multiplied in with matching (12) permutation 
symmetry. However, baryon wave functions exhibit the totally symmetric 
form (12)3+(23)1+(31)2, except for the color part, so that we need to 
symmetrize the spin invariants (12)3 that we have constructed so far. 

To make a totally symmetric spin-flavor wave function (without the color 
indices), we combine the (12) symmetric isospin $\vec \tau \otimes \vec \tau 
\phi_N$ with the symmetric (12) spin coupling, where 
$\phi_N$ is the isospin wave function of the nucleon. Similarly for the (12) 
antisymmetric combination, the mixed
antisymmetric isospin combination is given by $i\tau_2\otimes \phi_N$. Thus 
for the distinguishable (12)3 quark system we have the eight 
independent spin-isospin basis states shown in Table~\ref{123ndm}. 

Once we have the spinor invariants for the distinguishable quarks (12)3, we 
need to permute them according to (12)3+(23)1+(31)2. To simplify the basis 
further, we calculate the isospin matrix elements explicitly using the $uds$ 
basis, where in a proton, (12)3 denotes the quark configuration $uud$. 
Let us take $G_1$ as an example, 
\begin{eqnarray}
G_1=M\sum [\phi_1^{\dagger}i\tau_2\phi_2^{\dagger}
\phi_3\phi_N]_{isospin}\otimes[(C)_{12}\otimes (u_N)_{3}]_{spin}. 
\label{iso1}
\end{eqnarray}
Under permutation,
\begin{eqnarray}\nonumber
S_1 =M \sum [\phi_1^{\dagger}i\tau_2\phi_2\phi_3^{\dagger}\phi_N]_{isospin}
\otimes [(C)_{12}\otimes (u_N)_3]_{spin}+(23)1 +(31)2\\=[(23)1]_{spin}-
[(31)2]_{spin}. 
\label{iso2}
\end{eqnarray}
Dropping the subscript $spin$ in the above equation, the totally 
symmetric coupling $S_1$ corresponding to $G_1$  becomes 
\begin{eqnarray}
S_1 = (C)_{23} (u_N)_1 - (C)_{31} (u_N)_{2}. 
\label{iso4}
\end{eqnarray}
Similarly for a (12) symmetric ${\vec{\bf \tau}}_{12} \otimes 
{\vec{\bf \tau}}_3$ coupling in isospin space we find the isospin matrix 
elements -2,1,1 for (12)3, (23)1 and (31)2 couplings, respectively. Then the 
symmetrized spinor invariant $G_3$ becomes
\begin{eqnarray}
S_3 = -2M (\gamma^{\mu} C)_{12}\otimes (\gamma_5\gamma_{\mu} u_N)_3 + 
(23)1 +(31)2.
\label{iso5}
\end{eqnarray}

In order to rearrange the (23)1 and (31)2 terms in (12)3 order, we use the 
Fierz rearrangement Table~\ref{fierz} (see also Table I of ref.\cite{wap}) to 
get  
\begin{equation}
S_1 =  (23)1 -(31)2 =  {1\over 2} G_3-{1\over 4} G_7. 
\label{s1dm}
\end{equation} 
The final results are shown in Table~\ref{symndm}. From Table~\ref{symndm}, we 
can find out that there exist only three 
independent components, because there are five linear relationships between 
them, namely,
\begin{equation}
\begin{array}{lll}
S_1 & = & S_2 -S_4, \\
S_3 & = & 3 S_4,\\
S_5 & =& S_4-S_6,\\
S_7 & = & 12 S_2 - 6 S_4,\\
S_8 & =& 3S_2 - 2S_4+2S_6.
\end{array}
\label{lrn}
\end{equation}
Thus, we can choose $S_2, S_4, S_6$ as the three independent couplings for the  
nucleon wave function. 

The construction of spinor invariants for nucleon resonance 
states follows the same procdures except for replacing the total momentum 
spinor $u_N$ by a Rarita-Schwinger spinor for spin-$\frac{3}{2}$

\begin{eqnarray}
U_{\lambda}^{{1\over 2}^-,\mu} =
\sum_{m_1,m_2, m_1+m_2 =\lambda} C_{m_1 m_2 ~ {\lambda}} ^{1 ~~ {1\over 2} 
~~{1\over 2}}~{\hat\epsilon}_{m_1}^{\mu}u_{m_2}
\label{1520tmwf}
\end{eqnarray}

\begin{eqnarray}
U_{\lambda}^{{3\over 2}^-, \mu}=
\sum_{m_1,m_2, m_1+m_2 =\lambda} C_{m_1 m_2 ~{\lambda}} ^{1 ~~ {1\over 2} ~~ 
{3\over 2}}~{\hat\epsilon}_{m_1}^{\mu}u_{m_2},
\label{1535}
\end{eqnarray}
where $u_m(P)$ is a Dirac spinor, and the four mutually orthogonal vectors 
$P^{\mu}$, and the polarization vectors ${\hat\epsilon} _{\pm,0} ^{\mu}$ 
together form  basis of Minkowski space. The   
Rarita-Schwinger spinor has eight independent components because it 
satisfies the constraint $\gamma_{\mu}U_{\lambda}^{\frac{3}{2},\mu}=0$, which 
shows that there is no spin-$\frac{1}{2}$ contribution.
The orthogonality between the spinors is guaranteed by that of the 
Clebsch-Gordan coefficients.
The contraction between the relative momentum variables $s_{\mu}$ and 
$U_{1/2}^{1/2,{\mu}}$ gives 
\begin{eqnarray}
s_{\mu} U_{1\over 2}^{{1\over 2},\mu} = -{\sqrt {1\over
3}}\big ( s^zu_{\uparrow}+s^Ru_{\downarrow}\big )=-{\sqrt {1\over 3}} 
(\gamma_5\gamma\cdot s u_{\uparrow}).  
\label{suct}
\end{eqnarray}
This is consistent with the spinor invariants for $N^*(1535)$ constructed in
reference ~\cite{bkw}, where the total momentum wave function was written as 
$s_{\mu}u_{\lambda} $
and contracted with $\gamma^{\mu}$ instead of
$(s_{\mu} U_{1\over 2}^{{1\over 2},\mu})$ in our case.

For $N^*(1535),~N^*(1520)$ the invariants are constructed by replacing
the Dirac spinor by $\gamma_5(s_{\mu}U_{\lambda}^{({3\over 2},{1\over 2}),\mu})$
and are listed in Tables \ref{1535dm} and \ref{1520dm}. The extra 
$\gamma_5$ is necessary for the correct parity. 
The symmetrization among the three quarks follows the
same procedure as that for the nucleon and the results are listed in 
Table~\ref{sym1535dm} for the Dirac basis.

Because of the presence of
the relative momentum $s_{\mu}$ ($s_3=p_1-p_2$, etc.) the number of linear 
relations between the eight symmetrized states is reduced to three, namely
\begin{eqnarray}
2(S_1^{'}+S_2^{'})-S_7^{'}=0,\nonumber\\
 S_2^{'}+S_6^{'}-S_3^{'}+S_5^{'}-S_8^{'}=0,\nonumber \\
 4S_1^{'}+3S_3^{'}-S_4^{'}-S_7^{'}=0.
\label{lrn*}
\end{eqnarray}
This means that there are now five independent spinor components for 
$N^*(1535),~N^*(1520)$ each.

To summarize what we have accomplished so far, the eight independent 
relativistic couplings can be transformed into equivalent linear 
combinations of direct products of three Dirac spinors, which can be either 
$U_{\lambda}(P)$, the positive energy spinors or $V_{\lambda}(P)$, the 
negative energy spinors, 
where $\lambda$ means helicity. This is what is called Bargmann-Wigner basis 
in ref.~\cite{bkw} as compared to the Dirac matrix representation or basis. 
While the invariants $G_i$ of the Dirac basis can be easily interpretated as 
couplings between the three quarks and the nucleon, the invariants expressed 
in the BW basis have the advantage of being more easily permuted. Moreover,  
in Sect. IV, we shall use the BW representation as a means of transforming the 
spin-flavor wave functions from the instant to the LF form. 

\section{Three-Quark Nucleon States}
With $q^3-N$ couplings written in both Dirac and $U,~V$ forms, we are ready 
to generate the nucleon wave functions being used in light front dynamics. 
For simplicity we consider only the spin structure of the 
nucleon, ignore its isospin and treat quarks as distinguishable. 

The relativistic spinor basis is constructed from  
nonrelativistic Pauli spinors $\chi$. If we focus on the 
(12) quark pair, then the singlet pair state 
\begin{eqnarray}
|0,0\rangle\equiv\sum_{\lambda_1,\lambda_2}  C_{{\lambda}_1 ~{\lambda}_2 ~~{0}}
^{{1\over 2} ~~ {1\over 2}~~0} ~{\chi}_{\lambda_1}\otimes {\chi}_{\lambda_2}
\end{eqnarray}
the first step of the relativistic generalization is to replace Pauli spinors 
by Dirac spinors of positive energy in the instant form 
\begin{eqnarray}
|0,0\rangle^{(uu)} \equiv
\sum_{\lambda_1,\lambda_2}  
C_{{\lambda}_1 ~{\lambda}_2 ~~{0}} ^{{1\over 2} ~~ {1\over 2}~~0} 
~{u}_{\lambda}^{inst}\otimes u_{\lambda_2}^{inst}.  
\label{possinglet}
\end{eqnarray}
Upon transforming the instant to light-front spinors, we apply the first part 
of Eq.~(\ref{trf}) 
to get
\begin{eqnarray}
|0,0\rangle^{(uu)}=\sum_{\xi_1,\xi_2} \left[\sum_{\lambda_1,\lambda_2} 
\left(C_{{\lambda}_1 ~{\lambda}_2 ~~0}^{{1\over 2} ~~ {1\over 2}~~0} 
~{\cal R}^{(1)}_{\lambda_1\xi_1} {\cal R}^{(1)}_{\lambda_2\xi_2}\right)\right] 
~u_{\xi_1}^{LC} \otimes u_{\xi_2}^{LC}.  
\end{eqnarray}
Substituting Eq.~(\ref{rtrf}) for ${\cal R}_{\lambda\xi}^{(1)}$ we obtain 
\begin{equation}
|0,0\rangle^{(uu)} 
= {2m\over m+k^0} \sum_{\xi_1,\xi_2} \left[\bar u_{\xi_1}^{LC}
(\sum_{\lambda_1,\lambda_2} 
C_{{\lambda}_1 ~{\lambda}_2 ~~0} ^{{1\over 2} ~~ {1\over 2}~~0}
~U_{\lambda_1} \otimes U_{\lambda_2}) (\bar u_{\xi_2}^{LC})^T\right] 
~u_{\xi_1}^{LC} \otimes u_{\xi_2}^{LC}. 
\label{btold}
\end{equation}
From the conversion of the Dirac basis to the BW-basis in the 
previous Section III, we know that in the nucleon rest frame
\begin{eqnarray}
\sum_{\lambda_1,\lambda_2} C_{{\lambda}_1 ~{\lambda}_2 ~~0} ^{{1\over 2} ~~ 
{1\over 2}~~0}~U_{\lambda_1}\otimes U_{\lambda_2}=
{1+\gamma_0\over2\sqrt 2}\gamma_5 C
\end{eqnarray}
is valid. Therefore 
\begin{eqnarray}
|0,0\rangle^{(uu)} = {2m\over{\sqrt 2} (m+k^0)} \sum_{\xi_1,\xi_2} 
\left[\bar u_{\xi_1}^{LC} {1+\gamma_0\over2} \gamma_5C 
(\bar u_{\xi_2}^{LC})^T\right]~u_{\xi_1}^{LC} \otimes u_{\xi_2}^{LC}. 
\end{eqnarray}
This can be generalized to an arbitrary frame by boosting the nucleon to 
momentum P, so that we have
\begin{eqnarray}
|0,0\rangle^{(uu)} = {m\sqrt{2}\over m+k^0}\sum_{\xi_1,\xi_2} 
\left[\bar u_{\xi_1}^{LC} {\gamma\cdot P+M\over2M}\gamma_5 C 
(\bar u_{\xi_2}^{LC})^T\right]~u_{\xi_1}^{LC} \otimes u_{\xi_2}^{LC}.
\end{eqnarray}
Combining this quark pair state with the third quark we get precisely the 
three-quark state that is usually considered in applications of relativistic 
quark models. 

Now we are ready to extend the spin coupling method to quark pairs constructed  
via $V$-spinors, that is, there are similar new components starting from $z$ 
spinors in the instant form,  
\begin{equation}
\begin{array}{lll}
|0,0\rangle^{(zz)}&\equiv& \sum_{\lambda_1,\lambda_2}  
C_{{\lambda}_1 ~{\lambda}_2 ~~{0}} ^{{1\over 2} ~~ {1\over 2}~~0} 
~z_{\lambda}^{inst}\otimes z_{\lambda_2}^{inst}\\
&=& \sum_{\xi_1,\xi_2} \left[\sum_{\lambda_1,\lambda_2} 
C_{{\lambda}_1 ~{\lambda}_2 ~~0}^{{1\over 2} ~~ {1\over 2}~~0} 
~{\cal R}^{(4)}_{\lambda_1\xi_1}{\cal R}^{(4)}_{\lambda_2\xi_2}\right]
~u_{\xi_1}^{LC} \otimes u_{\xi_2}^{LC}\\
&=& -{2m\over k^0-m}\sum_{\xi_1,\xi_2} 
\left[\bar u_{\xi_1}^{LC} (\sum_{\lambda_1,\lambda_2} 
C_{{\lambda}_1 ~{\lambda}_2 ~~0} ^{{1\over 2} ~~ {1\over 2}~~0}
~V_{\lambda_1}\otimes V_{\lambda_2}) 
(\bar u_{\xi_2}^{LC})^T\right]~u_{\xi_1}^{LC} \otimes u_{\xi_2}^{LC}. 
\end{array}
\label{zznew}
\end{equation}
From Eq.~(\ref{zznew}) and Eq.~(\ref{btold}) we can form the linear combination
\begin{equation}
{{\sqrt 2}\over N^2}|0,0>^{(uu)}-{{\sqrt 2}\over N'^2}|0,0>^{(zz)}
=-\sum_{\xi_1,\xi_2}[\bar u_{\xi_1}^{LC}\gamma_5 C
(\bar u_{\xi_2}^{LC})^T]~u_{\xi_1}^{LC} \otimes u_{\xi_2}^{LC},
\end{equation}
where in the next to last line we recall (Eq.~\ref{g1})   
\begin{equation}
\begin{array}{lll}
-{1\over \sqrt 2}\gamma_5 C &=& {1\over\sqrt 2}
(U_{\uparrow}U_{\downarrow}-U_{\downarrow}U_{\uparrow}+
V_{\uparrow}V_{\downarrow}-V_{\downarrow}V_{\uparrow})\\
&=& \sum_{\lambda_1,\lambda_2}
C_{{\lambda}_1 ~{\lambda}_2 ~~0}^{{1\over 2} ~~ {1\over 2}~~0}
U_{\lambda_1}U_{\lambda_2}+\sum_{\lambda_1,\lambda_2}
C_{{\lambda}_1 ~{\lambda}_2 ~~0}^{{1\over 2} ~~ {1\over 2}~~0}
V_{\lambda_1}V_{\lambda_2}.
\label{g5C}
\end{array}
\end{equation}

Thus, from the linear combination of $|0,0\rangle^{(uu)}$ and 
$|0,0\rangle^{(zz)}$ we can construct the spinor invariant $G_2$ with positive 
spinors $u_1$ and $u_2$ in the Dirac representation. This means that, if we 
are to account for all eight invariants $G_i$ in Table~\ref{123ndm} as 
independent spinor components with the positive energy spinors as in 
light-front dynamics, then it is necessary to include the $z^{inst}$ spinors. 

With almost the same steps we find  
\begin{equation}
{{\sqrt 2}\over N^2}|0,0\rangle^{(uu)}+{{\sqrt 2}\over N'^2}|0,0\rangle^{(zz)}
= -\sum_{\xi_1,\xi_2}[\bar u_{\xi_1}^{LC}\gamma_0\gamma_5 C
(\bar u_{\xi_2}^{LC})^T]~u_{\xi_1}^{LC} \otimes u_{\xi_2}^{LC},
\end{equation}
which is the quark pair part of $G_6$ in the rest frame of the nucleon. There 
are similar expressions for all other spin invariants. 

We are now ready to complete the construction of relativistic, 
mixed-antisymmetric (MA) positive-energy three-quark spinor states.

From Eq.~(\ref{4matrix2}), we see that the $U$ and 
$V$ spinors come in when we transform the instant spinors to the light-front 
form. If we require the quark spinors to be of positive energy only, then 
there are only two ways to generate the $u_i^{LC}$
\begin{equation}
u_{\lambda}^{inst}=N\sum_{\xi}(\bar u_{\xi}^{LC}U_{\lambda}) 
u_{\lambda}^{LC},\quad
z_{\lambda}^{inst}=-N'\sum_{\xi}(\bar u_{\xi}^{LC}V_{\lambda}) u_{\lambda}^{LC}.
\label{meloshfull}
\end{equation}

Since in the $UV$ construction, the only allowed combinations are either 
three $U$s or two $V$s and one $U$ for parity reasons, in the instant form the 
quark spinors must be one of the following four combinations 
\begin{equation}
\begin{array}{lllll}
UUU &: & |MA,\lambda\rangle^{(uuu)} & \equiv &
\sum_{\lambda_3,\lambda}  
C_{0~{\lambda}_3 ~{\lambda}} ^{0~{1\over 2} ~~ {1\over 2}} 
(\sum_{\lambda_1,\lambda_2}  
C_{{\lambda}_1 ~{\lambda}_2 ~~{0}} ^{{1\over 2} ~~ {1\over 2}~~0} 
~{u}_{\lambda_1}^{inst}\otimes u_{\lambda_2}^{inst})
\otimes u_{\lambda_3}^{inst} \\
VVU &:& |MA,\lambda\rangle^{(zzu)}&\equiv&\sum_{\lambda_3,\lambda}  
C_{0~{\lambda}_3 ~{\lambda}} ^{0~{1\over 2} ~~ {1\over 2}} 
(\sum_{\lambda_1,\lambda_2}  
C_{{\lambda}_1 ~{\lambda}_2 ~~{0}} ^{{1\over 2} ~~ {1\over 2}~~0} 
~z_{\lambda_1}^{inst}\otimes z_{\lambda_2}^{inst})
\otimes u_{\lambda_3}^{inst} \\& & |MA,\lambda\rangle^{(zuz)} &\equiv &
\sum_{\lambda_3,\lambda}  
C_{0~{\lambda}_3 ~{\lambda}} ^{0~{1\over 2} ~~ {1\over 2}} 
(\sum_{\lambda_1,\lambda_2}  
C_{{\lambda}_1 ~{\lambda}_2 ~~{0}} ^{{1\over 2} ~~ {1\over 2}~~0} 
~z_{\lambda_1}^{inst}\otimes u_{\lambda_2}^{inst})
\otimes z_{\lambda_3}^{inst} \\& &|MA,\lambda\rangle^{(zzu)}&\equiv&
\sum_{\lambda_3,\lambda}  
C_{0~{\lambda}_3 ~{\lambda}} ^{0~{1\over 2} ~~ {1\over 2}} 
(\sum_{\lambda_1,\lambda_2}  
C_{{\lambda}_1 ~{\lambda}_2 ~~{0}} ^{{1\over 2} ~~ {1\over 2}~~0} 
~u_{\lambda_1}^{inst}\otimes z_{\lambda_2}^{inst})
\otimes z_{\lambda_3}^{inst}. 
\end{array}
\end{equation}

This gives a total of four states. We know their linear independence from 
their Dirac basis representations. It can be shown that linear 
combinations of states constructed from the above four states can generate 
$G_1$,~$G_2$,~$G_4$ and $G_6$. And similarly starting from the mixed-symmetric 
combination we can generate the remaining four $G_i$ couplings. These results 
display the explicit separation of the Dirac matrix forms of three-quark-nucleon 
couplings into states whose Dirac spinor content and spin couplings are 
manifest. In ref.~\cite{fbw} it is shown that different quark-nucleon wave 
functions characterized by various $G_i$ have important physical consequences. 
For example, the nonrelativistic quark model (NQM) gives the same fall-off with 
increasing momentum transfer for electric ($G^p_E$) and magnetic ($G^p_M$) 
form factors of the proton. The Melosh rotated ground state wave function of 
the NQM is given by $G_2+G_6$. However, wave functions of $G_2,$ $G_6$ and 
$G_2+G_6$ type generate different slopes of $G^p_E/G^p_M$ as functions of 
momentum transfer~\cite{fbw}, so that $G_2$ is in better agreement with recent 
data from Jefferson Laboratory~\cite{jl} than $G_2+G_6,$ or $G_6$. For the 
neutron charge form factor the agreement of $G_2$ with the data and 
disagreement of $G_2+G_6$ and $G_6$ are even more pronounced. 
   
The $UVV$ invariants differ from $UUU$ in that they can not be obtained from 
a well defined static limit via Melosh rotation. This explains why the $z,~w$ 
spinors do not have a well defined static limit which would circumvent this 
no-go theorem. In other words, the undefined static limit of the $z,~w$ spinors 
is the price one pays for the canonical Melosh construction of {\bf all} 
spin-flavor couplings of the Dirac basis. 

\section{Three-Quark States With Virtual Antiquarks}

In a typical electromagnetic form factor calculation the one-body quark 
current is sandwiched between three-quark wave functions of the 
nucleon~\cite{lcqm} that contain only $u$-quarks (see the $uuu-N$ vertex of 
Fig.1a). No intermediate $v$-quarks occur in the triangle Feynman diagram 
because the quark propagator  
contains only a $u$-spinor piece with energy denominator 
$p^--\frac{{\bf p}^2_{\perp}+m^2}{p^+}$ in addition to the instantaneous part 
$\frac{\gamma^+}{2p^+}$, and any $v$-spinor coupled to the virtual photon is 
eliminated in the Breit frame (by $q^+=0$). Nonetheless, the 
three-quark-nucleon and -baryon couplings may occur in Feynman diagrams 
sandwiched between $u$- and $v$-spinors. In this context, we expect that 
$v$-states, and the $vuu-N$ vertex of Fig.1b in particular, are generated by 
flux-tube breaking in QCD. In fact, in ref.~\cite{ki} the phenomenologically 
successful $^3$P$_0$ quark-pair creation model of hadron decays is generalized 
to the (color electric) flux tube breaking mechanism expected to occur in QCD 
at intermediate distances. Because it is conceptually simpler we keep 
the approximate $^3$P$_0$ quark pair creation vertex with its characteristic 
spin coupling~\cite{hjw} $\sim \bar u({\bf p})v(-{\bf p})\sim 
{\mathbf \sigma}\cdot {\bf p}$ in our discussion. The spin matrix elements (in 
spherical basis) are the Clebsch-Gordan coefficients that couple the quark-
antiquark spins to the triplet state. In hadronic decays the $^3$P$_0$ vertex 
introduces the $v$-spinor that converts a three-quark spin-flavor invariant of 
$uuu$ type to $vuu$ type. 

Now we are ready to extend the spin coupling method to quark pairs containing 
$v$-spinors and lift the restriction to positive energy spinors. Including two 
$v^{inst}$ spinors to maintain positive parity, and in analogy with 
Eq.~(\ref{possinglet}) we have
\begin{eqnarray}
|0,0\rangle^{(vv)}\equiv
\sum_{\lambda_1,\lambda_2}  
C_{{\lambda}_1 ~{\lambda}_2 ~~{0}} ^{{1\over 2} ~~ {1\over 2}~~0} 
~{v}_{\lambda}^{inst}\otimes v_{\lambda_2}^{inst}.  
\label{vv}
\end{eqnarray}

Applying the Melosh rotation ${\cal R}^{(2)}_{\lambda\xi}$~(Eq.~\ref{rtrf}), we 
obtain from Eq.~\ref{vv}
\begin{equation}
\begin{array}{lll}
|0,0\rangle^{(vv)} &=& \sum_{\lambda_1,\lambda_2}  
C_{\lambda_1 ~\lambda_2 ~~0}^{{1\over 2} ~~ {1\over 2}~~0}~\sum_{\xi_1,\xi_2} 
({\cal R}_{\lambda_1\xi_1}^{(2)} v_{\xi_1}^{LC})\otimes
({\cal R}_{\lambda_2\xi_2}^{(2)} v_{\xi_2}^{LC})\\ 
&=& -N^2 \sum_{\xi_1,\xi_2} \left[\sum_{\lambda_1,\lambda_2} 
C_{{\lambda}_1 ~{\lambda}_2 ~~0}^{{1\over 2} ~~ {1\over 2}~~0} 
~(\bar v_{\xi_1}^{LC}V_{\lambda_1})
(\bar v_{\xi_2}^{LC}V_{\lambda_2})\right]~v_{\xi_1}^{LC} \otimes 
v_{\xi_2}^{LC}\\
&=& -{2m\over m+k^0}\sum_{\xi_1,\xi_2} \left[\bar v_{\xi_1}^{LC} 
(\sum_{\lambda_1,\lambda_2} C_{{\lambda}_1 ~{\lambda}_2 ~~0}^{{1\over 2} ~~ 
{1\over 2}~~0}~V_{\lambda_1}\otimes V_{\lambda_2}) (\bar v_{\xi_2}^{LC})^T
\right] ~v_{\xi_1}^{LC} \otimes v_{\xi_2}^{LC}. 
\end{array}
\label{vvn}
\end{equation}
From the conversion in Section III, we recall that  
\begin{equation}
\begin{array}{lll}
-\sum_{\lambda_1,\lambda_2}
C_{{\lambda}_1 ~{\lambda}_2 ~~0}^{{1\over 2} ~~ {1\over 2}~~0}
V_{\lambda_1}V_{\lambda_2} &=&
\sum_{\lambda_1,\lambda_2}
C_{{\lambda}_1 ~{\lambda}_2 ~~0}^{{1\over 2} ~~ {1\over 2}~~0}
U_{\lambda_1}U_{\lambda_2}-{1\over\sqrt 2}\gamma_5 C\\
&=& {1\over\sqrt 2}({1+\gamma_0\over 2}\gamma_5 C-\gamma_5 C)
= {1\over\sqrt 2}({\gamma_0-1\over 2}\gamma_5 C). 
\label{vvo}
\end{array}
\end{equation}
Substituting Eq.~(\ref{vvo}) into Eq.~(\ref{vvn}) we get
\begin{eqnarray}
|0,0\rangle^{(vv)}={m\sqrt{2}\over m+k^0}\sum_{\xi_1,\xi_2} 
\left[\bar v_{\xi_1}^{LC}{\gamma_0-1\over 2}\gamma_5 C(\bar v_{\xi_2}^{LC})^T
\right]~v_{\xi_1}^{LC} \otimes v_{\xi_2}^{LC}.
\label{vvnew2}
\end{eqnarray}
Again, Eq.~(\ref{vvnew2}) can be generalized to a frame where the nucleon has 
momentum $P$, which gives
\begin{eqnarray}
|0,0\rangle^{(vv)}={m\sqrt{2}\over m+k^0}\sum_{\xi_1,\xi_2} 
\left[\bar v_{\xi_1}^{LC}{\gamma\cdot P-M\over 2M}\gamma_5 C 
(\bar v_{\xi_2}^{LC})^T\right]~v_{\xi_1}^{LC} \otimes v_{\xi_2}^{LC}.
\label{vvnew3}
\end{eqnarray}
From Eq.~\ref{vvnew3} we can see that, by applying the Melosh transformation 
to the $v^{inst}$ spinors, a new component of the Dirac representation is 
generated. All others are constructed similarly. 

For completeness we mention three-quark states of spin $\frac{1}{2}$ and 
negative parity with a single $v$ spinor that we expect to play a role in 
sea-quark Fock states 
\begin{equation}
\begin{array}{lllll}
vuu &:& |MA,\lambda\rangle^{(wuu)}&\equiv&\sum_{\lambda_3,\lambda}  
C_{0~{\lambda}_3 ~{\lambda}} ^{0~{1\over 2} ~~ {1\over 2}} 
(\sum_{\lambda_1,\lambda_2}  
C_{{\lambda}_1 ~{\lambda}_2 ~~{0}} ^{{1\over 2} ~~ {1\over 2}~~0} 
~w_{\lambda_1}^{inst}\otimes u_{\lambda_2}^{inst})
\otimes u_{\lambda_3}^{inst} \\& & |MA,\lambda\rangle^{(uwu)} &\equiv &
\sum_{\lambda_3,\lambda}  
C_{0~{\lambda}_3 ~{\lambda}} ^{0~{1\over 2} ~~ {1\over 2}} 
(\sum_{\lambda_1,\lambda_2}  
C_{{\lambda_1} ~{\lambda}_2 ~~{0}} ^{{1\over 2} ~~ {1\over 2}~~0} 
~u_{\lambda_1}^{inst}\otimes w_{\lambda_2}^{inst})
\otimes u_{\lambda_3}^{inst} \\& &|MA,\lambda\rangle^{(uuw)}&\equiv&
\sum_{\lambda_3,\lambda}  
C_{0~{\lambda}_3 ~{\lambda}} ^{0~{1\over 2} ~~ {1\over 2}} 
(\sum_{\lambda_1,\lambda_2}  
C_{{\lambda}_1 ~{\lambda}_2 ~~{0}} ^{{1\over 2} ~~ {1\over 2}~~0} 
~u_{\lambda_1}^{inst}\otimes u_{\lambda_2}^{inst})
\otimes w_{\lambda_3}^{inst}. 
\end{array}
\end{equation}
There are also the corresponding mixed symmetric $vuu$ spin invariants, 
which result from replacing the Clebsch-Gordan coefficients for pair spin $0$ 
by those for pair spin $1$. 

A flux tube breaking in 
the nucleon, then, involves a quark pair creation event from the vacuum and 
generates an intermediate $v$-state, which may be converted by a valence gluon 
as in Fig.2a, or by a Goldstone boson in effective chiral field theory as in 
Fig.2b, into a $u$-spinor. Thus, these diagrams represent transition amplitudes 
from a three-quark Fock state to a three-quark-gluon or a three-quark-Goldstone 
boson Fock state mediated by quark pair creation. As suggested by chiral 
quark models, these Fock states are expected to contribute to the neutron 
charge form factor in particular, which relativistic quark model can not 
explain on the basis of the canonical ($UUU$) three-quark wave function alone.  
However, in scalar coupling ($G_2$ of Table~\ref{123ndm}) the neutron charge 
form factor description improves~\cite{fbw}. The $vuu$ states also enter and 
are probed by time-like weak processes discussed in ref.~\cite{dgns}. 

A key feature of the quark-nucleon coupling involving an antiquark 
is that the vertex function $\Lambda$  is in general no longer directly 
related to the radial wave function. This point has been emphasized 
in recent analyses~\cite{tm,k} of skewed parton distributions  
of deeply virtual Compton scattering (DVCS) from the proton. When the leading 
twist handbag diagram of DVC is integrated over the longitudinal momentum 
fraction $x,$ the integrand of the covariant triangle diagram results. In 
light-cone time-ordered perturbation theory the latter splits up into the 
standard form factor result involving the initial and final light-cone wave 
functions of the proton and the diagram shown in Fig.~3, where the photon 
momentum $q^+>0$ can no longer be chosen to vanish in DVCS. The 
antiquark-$q^2$-nucleon vertex function $\Lambda$ involves $k^{+}-q^+<0.$   
Using the Bethe-Salpeter equation for the proton projected to the null-plane, 
where it becomes the Weinberg equation, the negative momentum fraction may be  
shifted into the kernel $V$ of the equation of motion, 
whereas the proton light-cone wave function $\psi$ retains positive 
plus-momentum. In our case, using the $^3P_0$ pair creation amplitude 
$g_p \bar v_1 u_1,$ the proton spin-flavor wave function 
$\bar u_2\gamma_5C\bar u_3^T(\bar u_1u_N)-\bar u_1\gamma_5C\bar u_3^T
(\bar u_2u_N)$ is converted through the $u_1\bar u_1$ spin sum to the 
spin-flavor vertex structure $$\bar u_2\gamma_5C\bar u_3^T(\bar v_1(q-k)u_N)
-\bar v_1(q-k) \gamma_5C\bar u_3^T(\bar u_2u_N),$$ 
where the $v$-spinor has the proper positive plus-component as a consequence 
of the pair creation from the vacuum. The vertex function 
$\Lambda\sim g_p R(M^2_0)(m_N^2-M_0^2)$ remains connected to the radial 
light-cone momentum wave function $R$ of the final proton state.       
     
\section{Discussion And Conclusions}   

We have constructed relativistic three-quark states for the nucleon and 
several N$^*$s in the Dirac representation and compared with the 
Bargmann-Wigner basis by systematically including all 
dynamically accessible matrix elements ($\bar uV$), ($\bar vU$), and 
($\bar vV$). Our direct and transparent construction of the Dirac basis 
states makes its equivalence with the Bargmann-Wigner basis manifest, thus 
avoiding the evaluation of many-body overlap matrix elements of 
ref.~\cite{bkw}. We have shown that in light front dynamics the nonstatic 
quark-baryon couplings (wave functions with zero static limit) can be 
constructed via unitary transformations from the instant form as well. 
Therefore, in a Lorentz covariant framework, they form part of a Hilbert space 
and should be treated on equal footing with the Melosh rotated nonrelativistic 
states. We also have compiled all linear relations among the symmetrized 
quark-nucleon couplings, as well as those for the N$^*$(1520) and N$^*$(1535) 
nucleon resonance states, from which the number of independent basis states 
follows. The Melosh transformations facilitate the construction of states with 
one or two antiquarks which are ingredients in transition amplitudes from 
three-quark to three-quark-gluon or three-quark-Goldstone boson Fock states.  
   
\section{Acknowledgement}
We are grateful for the support of the INPP at UVa. 

\begin{table}
\begin{center}
\begin{tabular}[2pt] {ll}
\\ \hline\hline
$G_1=$ \qquad
 $M(i{\tau}_2 C)_{12}\otimes (\gamma_5       
u_{\lambda})_3$
\\ \hline
$G_2=$  \qquad
 $M(i{\tau}_2\gamma_5 C)_{12}\otimes (u_{\lambda})_3$  \\ \hline
$G_3=$  \qquad
$M(\gamma^{\mu}{\vec \tau}i\tau_2 C)_{12}\otimes \vec {\tau} (\gamma_5
\gamma_{\mu}u_{\lambda})_3$
\\ \hline
$G_4=$  \qquad
 $M(i{\tau}_2\gamma^{\mu}\gamma_5 C)_{12}\otimes
     (\gamma_{\mu}u_{\lambda})_3$ \\ \hline
$G_5=$ \qquad
 $\gamma \cdot Pi\tau_2{\vec \tau} C \otimes 
  (\gamma_5{\vec \tau}u_{\lambda})_3$ \\ \hline
$G_6=$ \qquad
 $(i{\tau}_2\gamma \cdot P\gamma_5 C)_{12}\otimes
            (u_{\lambda})_3$ \\ \hline
$G_7=$ \qquad
$M(\sigma^{\mu\nu}{\vec \tau}i\tau_2 C)_{12}\otimes(\gamma_5 
\sigma_{\mu\nu}{\vec \tau}u_{\lambda})_3$ \\ \hline
$G_8=$ \qquad
$i(\sigma^{\mu\nu}P_{\nu}{\vec \tau}i\tau_2 C)_{12}\otimes ({\vec \tau}
\gamma_5\gamma_{\mu}u_{\lambda})_3$ \\ \hline
\end{tabular}
\end{center}
Note that in each of the bases $u_{\lambda}$ is understood as containing both 
the isospin and spin components of the nucleon. $C=i\gamma^2\gamma^0$ is the 
charge conjugation matrix.
\caption{\bf {Spinor invariants of the nucleon in (12)3 coupling of the Dirac 
basis}} 
\label{123ndm}
\end{table}

\begin{table}
\begin{center}
\begin{tabular}[2pt] {ll}
$G_1=$& 
 $M(UV+VU)\otimes(-\uparrow\downarrow+\downarrow\uparrow)\otimes V^{\uparrow}$
\\ \hline
$G_2=$ & 
 $M(UU+VV)\otimes(-\uparrow\downarrow+\downarrow\uparrow)\otimes U^{\uparrow}$
\\ \hline
$G_3=$ & 
 $-M[(UV-VU)\otimes (\uparrow\downarrow-\downarrow\uparrow)\otimes V^{\uparrow}
  -2(UU-VV)\otimes (\uparrow\uparrow)\otimes U^{\downarrow}$\\
 &$ +(UU-VV)\otimes (\uparrow\downarrow+\downarrow\uparrow)\otimes U^{\uparrow}]$
\\ \hline
$G_4=$ & 
 $-M[(UU-VV)\otimes(\uparrow\downarrow-\downarrow\uparrow)\otimes U^{\uparrow}
  -2(UV-VU)\otimes(\uparrow\uparrow)\otimes V^{\downarrow}$\\
& $ +(UV-VU)\otimes (\uparrow\downarrow+\downarrow\uparrow)\otimes V^{\uparrow}]$
\\ \hline   
$G_5=$ & 
 $-M(UV-VU)\otimes (\uparrow\downarrow-\downarrow\uparrow)\otimes V^{\uparrow}$
\\ \hline
$G_6=$&
 $-M(UU-VV)\otimes(\uparrow\downarrow-\downarrow\uparrow)\otimes U^{\uparrow}$
\\ \hline
$G_7=$ &
 $-2M[(UU+VV)\otimes(\uparrow\downarrow+\downarrow\uparrow)\otimes U^{\uparrow}
  -2(UU+VV)\otimes(\uparrow\uparrow)\otimes U^{\downarrow}$\\
& $ +(UV+VU)\otimes (\uparrow\downarrow+\downarrow\uparrow)\otimes V^{\uparrow}
  -2(UV+VU)\otimes(\uparrow\uparrow)\otimes V^{\downarrow}]$
\\ \hline
$G_8=$& 
 $-M[-2(UU+VV)\otimes(\uparrow\uparrow)\otimes U^{\downarrow}
  +(UU+VV)\otimes(\uparrow\downarrow+\downarrow\uparrow)\otimes U^{\uparrow}]$\\
 \hline
\end{tabular}
\end{center}
\caption{\bf {Conversion Table from DM Basis to BW Basis}}
\label{dmbw}
\end{table}

\begin{table}
\begin{center}
\begin{tabular}[9pt] {cccccc} \hline
  & $J_S$ & $J_V$ & $J_T$ & $J_A$ & $J_P$
\\ \hline\hline
$J_S'$ & 1/4 & 1/4 & 1/8 & -1/4 & 1/4
\\ \hline
$J_V'$ & 1 & -1/2 & 0 & -1/2 & -1
\\ \hline
$J_T'$ & 3 & 0 & -1/2 & 0 & 3
\\ \hline
$J_A'$ & -1 & -1/2 & 0 & -1/2 & 1
\\ \hline
$J_P'$ & 1/4 & -1/4 & 1/8 & 1/4 & 1/4
\\ \hline
\end{tabular}
\end{center}
Example:  $J_V'=J_S-{1\over 2}J_V-{1\over 2}J_A-J_P$ \\
Definitions:\\
$J_S=(\bar u_1u_2)(\bar u_3u_N)$\\ $J_S'=(\bar u_3u_2)(\bar u_1u_N)$\\
$J_V=(\bar u_1\gamma_{\mu}u_2)(\bar u_3\gamma^{\mu}u_N)$\\ 
$J_V'=(\bar u_3\gamma_{\mu}u_2)(\bar u_1\gamma^{\mu}u_N)$\\
$J_T=(\bar u_1\sigma_{\mu\nu}u_2)(\bar u_3\sigma^{\mu\nu}u_N)$\\ 
$J_T'=(\bar u_3\sigma_{\mu\nu}u_2)(\bar u_1\sigma^{\mu\nu}u_N)$\\
$J_A=(\bar u_1\gamma_5\gamma_{\mu}u_2)(\bar u_3\gamma^5\gamma^{\mu}u_N)$\\ 
$J_A'=(\bar u_3\gamma_5\gamma_{\mu}u_2)(\bar u_1\gamma^5\gamma^{\mu}u_N)$\\
$J_P=(\bar u_1\gamma_5u_2)(\bar u_3\gamma^5u_N)$\\
$J_P'=(\bar u_3\gamma_5u_2)(\bar u_1\gamma^5u_N)$

\caption{{\bf Fierz Transformation Table}}
\label{fierz}
\end{table}

\begin{table}
\begin{center}
\begin{tabular}[9pt] {cccccccccc} \hline
  & $G_1$ & $G_2$ & $G_3$ & $G_4$ & $G_5$
  & $G_6$ & $G_7$ & $G_8$ & 
\\ \hline\hline
$S_1$ & 0 & 0 & 1/2 & 0 & 0 & 0 & -1/4 & 0
\\ \hline
$S_2$ & 0 & 0 & -1/2 & 0 & 0 & 0 & -1/4 & 0
\\ \hline
$S_3$ & 0 & 0 & -3 & 0 & 0 & 0 & 0 & 0 
\\ \hline
$S_4$ & 0 & 0 & -1 & 0 & 0 & 0 & 0 & 0
\\ \hline
$S_5$ & 0 & 0 & -1/2 & 0 & -1 & 0 & -1/4 & 1 
\\ \hline
$S_6$ & 0 & 0 & -1/2 & 0 & 1 & 0 & 1/4 & -1
\\ \hline
$S_7$ & 0 & 0 & 0 & 0 & 0 & 0 & -3 & 0 
\\ \hline
$S_8$ & 0 & 0 & -1/2 & 0 & 2 & 0 & -1/4 & -2
\\ \hline
\end{tabular}
\end{center}
\caption{{\bf {Symmetrized Spinor Invariants for $N$}}}
\label{symndm}
\end{table}

\begin{table}
\begin{center}
\begin{tabular}[2pt] {ll}
\\ \hline\hline
$G_1=$ \quad
$M({\vec \tau}i{\tau}_2 C)_{12}\otimes({\vec \tau}\gamma_5 
      (s_{\mu}\cdot U_{\lambda}^{{1\over 2},\mu}))_3$
\\ \hline
$G_2=$  \qquad
 $M({\vec \tau}i{\tau}_2\gamma_5 C)_{12}\otimes({\vec \tau}         
   (s_{\mu} \cdot U_{\lambda}^{{1\over 2},\mu}))_3$  \\ \hline
$G_3=$  \qquad
 $M(\gamma^{\mu}i\tau_2 C)_{12}\otimes
(\gamma_5\gamma_{\mu}(s_{\mu}\cdot U_{\lambda}^{{1\over 2},\mu}))_3$
\\ \hline
$G_4=$  \qquad
 $M({\vec \tau}i{\tau}_2\gamma^{\mu}\gamma^5 C)_{12} \otimes({\vec \tau}
\gamma_{\mu}(s_{\mu}\cdot U_{\lambda}^{{1\over
2},\mu}))_3$\\ \hline 
$G_5=$ \qquad
$(\gamma \cdot Pi\tau_2 C)_{12}\otimes(\gamma_5
(s_{\mu} \cdot U_{\lambda}^{{1\over 2},\mu}))_3$
 \\ \hline
$G_6=$ \qquad  
  $({\vec \tau}i{\tau}_2\gamma \cdot P\gamma_5 C)_{12}\otimes ({\vec \tau}
  (s_{\mu}\cdot U_{\lambda}^{{1\over 2},\mu}))_3$
\\ \hline
$G_7=$ \qquad
$M(\sigma^{\mu\nu}i\tau_2 C)_{12}\otimes(\gamma_5\sigma_{\mu\nu}(s_{\mu}
\cdot U_{\lambda}^{{1\over
2},\mu}))_3$
\\ \hline
$G_8=$ \qquad
$(i\sigma^{\mu\nu}P_{\nu}i\tau_2 C)_{12}\otimes (\gamma_5\gamma_{\mu}
(s_{\mu}\cdot U_{\lambda}^{{1\over 2},\mu}))_3$\\ \hline
Note: $U_{\lambda}^{{1\over 2},\mu} =
C_{m_1m_2\lambda}^{1~{1\over 2}~{1\over 2}}{\hat \epsilon_{m_1}}u_{m_2}=$
$-\sqrt {1\over 3} $
$\hat \epsilon_0^{\mu}u_{1\over2}+\sqrt {2\over 3} \hat \epsilon_{+1}^{\mu}
u_{-{1\over2}}$\\ \hline
\end{tabular}
\end{center}
\caption{\bf {Spinor Invariants for $N^*(1535)$ in (12)3 coupling of the Dirac 
basis}}
\label{1535dm}
\end{table}

\begin{table} 
\begin{center} 
\begin{tabular}[2pt] {ll} 
\\ \hline\hline 
$G_1=$ \quad 
$M({\vec \tau}i{\tau}_2 C)_{12}\otimes({\vec \tau}\gamma_5  
      (s_{\mu}\cdot U_{\lambda}^{{3\over 2},\mu})_3)$ 
\\ \hline 
$G_2=$  \qquad 
 $M({\vec \tau}i{\tau}_2\gamma_5 C)_{12}\otimes({\vec \tau}          
   (s_{\mu} \cdot U_{\lambda}^{{3\over 2},\mu}))_3$  \\ \hline 
$G_3=$  \qquad 
 $M(\gamma^{\mu}i\tau_2 C)_{12}\otimes 
(u_3\gamma_5\gamma_{\mu}(s_{\mu}\cdot U_{\lambda}^{{3\over 2},\mu}))_3$ 
\\ \hline 
$G_4=$  \qquad 
 $M({\vec \tau}i{\tau}_2\gamma^{\mu}\gamma^5 C)_{12}\otimes({\vec \tau}
\gamma_{\mu}(s_{\mu}\cdot U_{\lambda}^{{3\over 2},\mu}))_3$ 
\\ \hline  
$G_5=$ \qquad 
$(\gamma \cdot Pi\tau_2 C\bar u_2^T)_{12}\otimes (\gamma_5 
(s_{\mu} \cdot U_{\lambda}^{{3\over 2},\mu}))_3$ 
 \\ \hline 
$G_6=$ \qquad   
  $({\vec \tau}i{\tau}_2\gamma \cdot P\gamma_5 C)_{12}\otimes ({\bf \tau}
  (s_{\mu}\cdot U_{\lambda}^{{3\over 2},\mu}))_3$ 
\\ \hline 
$G_7=$ \qquad 
$M(\sigma^{\mu\nu}i\tau_2 C)_{12}\otimes(\gamma_5\sigma_{\mu\nu}(s_{\mu}
\cdot U_{\lambda}^{{3\over 
2},\mu}))_3$ 
\\ \hline 
$G_8=$ \qquad 
$i(\sigma^{\mu\nu}P_{\nu}i\tau_2 C)_{12}\otimes(\gamma_5\gamma_{\mu}
(s_{\mu}\cdot U_{\lambda}^{{3\over 2},\mu}))_3$ 
\\ \hline 
Note: $U_{\lambda}^{{3\over 2},\mu} = 
C_{m_1m_2\lambda}^{1~{1\over 2}~{3\over 2}}{\hat \epsilon_{m_1}}u_{m_2},
$\quad
$U_{3\over 2}^{{3\over 2},\mu} = {\hat \epsilon_{1}}u_{1\over 2},$\quad
$U_{1\over 2}^{{3\over 2},\mu} = \sqrt {2\over 3}  
\hat \epsilon_0^{\mu}u_{1\over2}+\sqrt {1\over 3} \hat
\epsilon_{+1}^{\mu}u_{-{1\over2}}$.\\ \hline 
\end{tabular} 
\end{center} 
\caption{\bf {Spinor Invariants for $N^*(1520)$ in (12)3 coupling of the Dirac 
basis}}
\label{1520dm}
\end{table} 

\begin{table}
\begin{center}
\begin{tabular}[9pt] {cccccccccc} \hline
  & $G_1$ & $G_2$ & $G_3^*$ & $G_4$ & $G_5^*$
  & $G_6$ & $G_7^*$ & $G_8$ & 
\\ \hline\hline
$S_1$ & -7/4 & 1/4 & 1/4 & -1/4 & 0 & 0 & 1/8 & 0
\\ \hline
$S_2$ & 1/4 & -7/4 & -1/4 & 1/4 & 0 & 0 & 1/8 & 0
\\ \hline
$S_3$ & -1 & 1 & 1/2 & 1/2 & 0 & 0 & 0 & 0 
\\ \hline
$S_4$ & -1 & 1 & -1/2 & -5/2 & 0 & 0 & 0 & 0
\\ \hline
$S_5$ & 1/4 & -1/4 & -1/4 & -1/4 & -1/2 & 1/2 & 1/8 & 1/2 
\\ \hline
$S_6$ & -1/4 & 1/4 & -1/4 & -1/4 & -1/2 & -3/2 & -1/8 & -1/2
\\ \hline
$S_7$ & -3 & -3 & 0 & 0 & 0 & 0 & 1/2 & 0 
\\ \hline
$S_8$ & 3/4 & 3/4 & 1/4 & -1/4 & 1 & 1 & -1/8 & 0
\\ \hline
&
&$G_i \sim s_{3,\mu}\cdot U_{1\over 2}^{{1\over 2},\mu}$, where 
$s_3 = p_1-p_2$\\ \hline
&
&$G_i^* \sim s_{m,\mu}\cdot U_{1\over 2}^{{1\over 2},\mu}$, where 
$s_m = 2p_3 - p_1 -p_2 = s_2 -s_1$
\\ \hline 
\end{tabular}
\end{center}
\caption{{\bf {Symmetrized Spinor Invariants for $N^*(1535)$} in DM Basis}}
\label{sym1535dm}
\end{table}

\begin{figure}[h]
\vglue 2in
\centerline{\psfig{figure=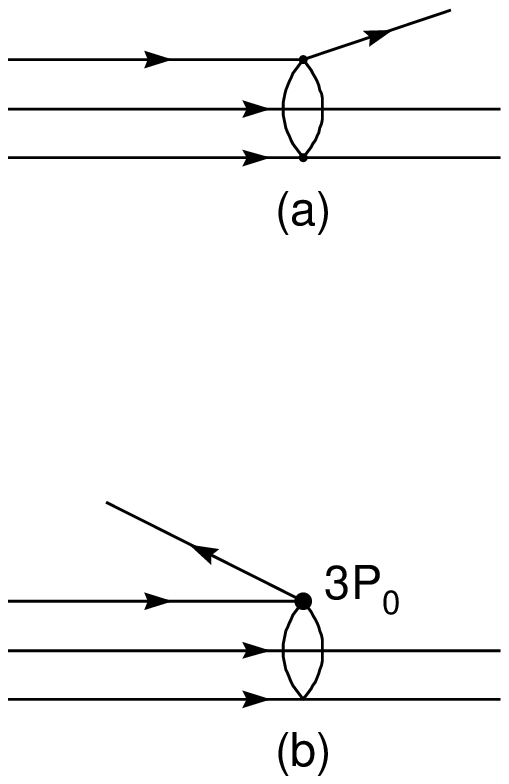,width=4in}}
\vglue 2in
\caption{(a) Three-quark-nucleon vertex: uuu; (b) uuv via quark pair creation}
\end{figure}

\begin{figure}[h]
\vglue 2in
\centerline{\psfig{figure=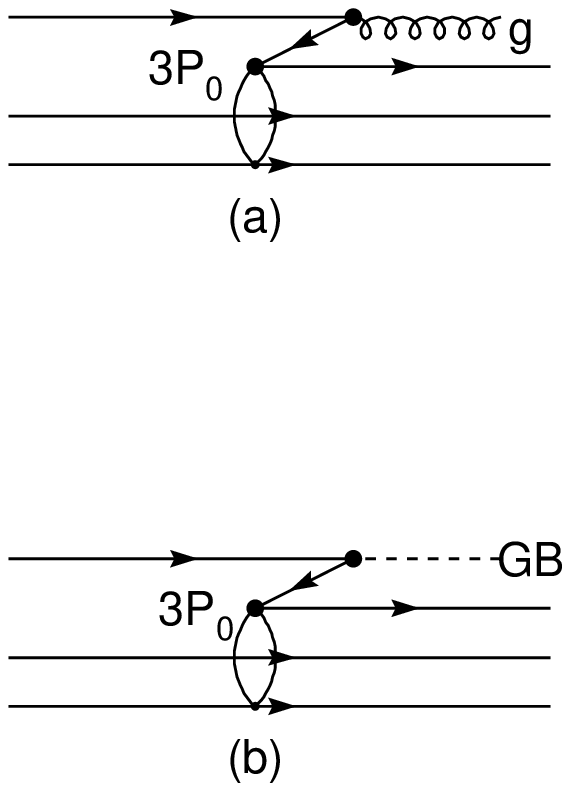,width=4in}}
\vglue 2in
\caption{(a) Three-quark to three-quark-gluon Fock state transition amplitude; 
(b) three-quark to three-quark-Goldstone boson Fock state transition amplitude}
\end{figure}

\begin{figure}[h]
\vglue 2in
\centerline{\psfig{figure=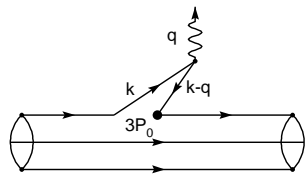,width=4in}}
\vglue 2in
\caption{(a) Triangle Z-diagram}
\end{figure}

\end{document}